# The correlation between fragility, density and atomic interaction in glass-forming liquids

Lijin Wang[1], Pengfei Guan[1*], and W. H. Wang[2]

[1]Beijing Computational Science Research Center, Beijing, 100193, P. R. China.

[2]Institute of Physics, Chinese Academy of Sciences, Beijing 100190, P. R. China

*Corresponding E-mail: pguan@csrc.ac.cn

The fragility that controls the temperature-dependent viscous properties of liquids as the glass transition is approached, in various glass-forming liquids with different softness of the repulsive part of atomic interactions at different densities is investigated by molecular dynamic simulations. We show the landscape of fragility in purely repulsive systems can be separated into three regions denoted as **R<sub>I</sub>**, **R<sub>II</sub>** and **R<sub>III</sub>**, respectively, with qualitatively disparate dynamic behaviors: **R<sub>I</sub>** which can be described by 'softness makes strong glasses', **R<sub>II</sub>** where fragility is independent of softness and can only be tuned by density, and **R<sub>III</sub>** with constant fragility, suggesting that density plays an unexpected role for understanding the repulsive softness dependence of fragility. What more important is that we unify the long-standing inconsistence with respect to the repulsive softness dependence of fragility by observing that a glass former can be tuned more fragile if nonperturbative attraction is added into it. Moreover, we find that the vastly dissimilar influences of attractive interaction on fragility could be estimated from the structural properties of related zero-temperature glasses.

## I. INTRODUCTION

Understanding the nature of glass transition is still a major open question in condensed matter physics, [1-6] though great efforts have been made over the years. An important aspect of this question is elucidating how the structural relaxation time $\tau$ of a glass-forming liquid increases significantly on cooling and drives the system towards its glassy state at some temperature $T_g$. The kinetic fragility index[7-9] $m$ which is usually defined by $m = \left.\frac{d \log \tau}{d \left(\frac{T_g}{T}\right)}\right|_{T=T_g}$, characterizes how rapidly the $log(\tau)$ changes with $T_g/T$ at $T = T_g$ and presents the singular dynamic behaviours of different glass-forming liquids. $T_g$ is the glass transition temperature below which the material has become so viscous that its typical relaxation time scale (of the order of 100s)[5] exceeds the measurable time window. The fragility $m$ exhibits great diversity over a wide swath of glass formers from colloids[3] to atoms[5], and the liquids can be defined as strong or fragile liquids with small or large $m$, respectively. Despite the fact that numerous empirical correlations between fragility and other dynamical, [10,11] thermo-dynamical, [9,10,12] vibrational[13] or mechanical[14] properties have been observed in different glass formers, the physical origin of fragility, deemed to be one of the key issues likely leading to the ultimate understanding of glass transition, remains unresolved.

Recently, fragility has been investigated in a series of experimental and theoretical studies[15-21] by taking into consideration the details of constituent interatomic potentials which underpin many properties of glasses and glass formers. [22-25] Apparent correlations between fragility and specific properties of interaction potential have been observed. However, some of them beyond our expectations have so far proved inconclusive or contrary. One crucial but inconclusive question up to now is how softness of repulsion affects fragility. In the colloidal experiment, [15] it is claimed that softness makes strong glasses, while different conclusions have also been drawn, e.g. softness increases fragility in modified Lennard-Jonnes systems, [16-19] and fragility is independent of softness in inverse power law potential systems. [20]

In this paper, a series of glass forming systems composing of particles with various densities, interacting via modified Lennard-Jones potentials with different steepness of the repulsive interaction, were studied to investigate the correlation between the fragility, density and atomic interaction. We unified the long-standing inconsistence[15-20] with respect to the repulsive softness dependence of fragility. It is found that the landscape of fragility in purely repulsive systems can be separated into three regions with qualitatively disparate dynamic behaviours. Furthermore, the influence of attractive interaction on fragility can be estimated from the structural properties of related zero-temperature glass.

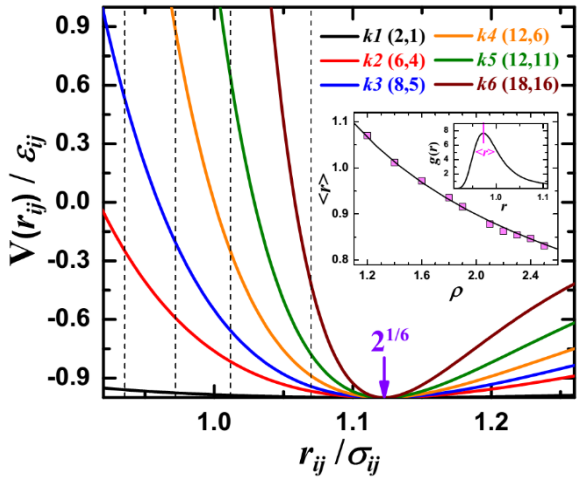

FIG. 1. Schematic illustration of the scaled Lennard-Jones potential $V(r_{ij})/\epsilon_{ij}$. Inset: Density $\rho$ dependence of the position $\langle r \rangle$, as illustrated in the sub-inset, of the first peak of pair distribution function of A particles $g_{AA}(r)$ of inherent structures for model $k4_{rlj}$. The solid line in the inset is the fit according to $\langle r \rangle \sim \rho^{-1/3}$. The vertical dashed lines from right to left mark $\langle r \rangle$ at $\rho = 1.2, 1.4, 1.6$ and $1.8$, respectively. The purple arrow locates the minimum of $V(r_{ij})/\epsilon_{ij}$.

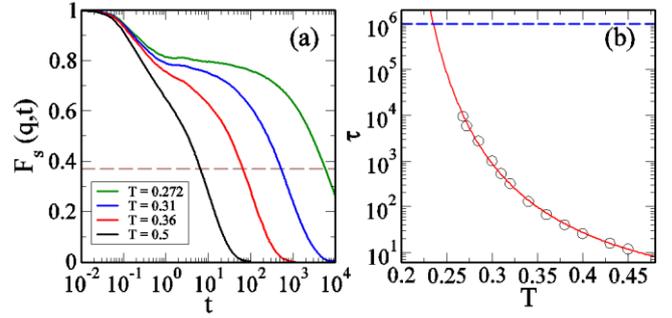

FIG. 2. (a): Temperature $T$ dependence of the self-intermediate scattering function $F_s(q,t)$ of glass forming liquids at density, $\rho = 1.2$ for model $k4_{rlj}$ without attraction. Relaxation time $\tau$ at each $T$ is determined from $F_s(q,t) = e^{-1}$ which is marked with horizontal dashed line. (b): $T$ dependence of $\tau$. The solid line is fit to VFT function. The horizontal dashed line marks $\tau = 10^6$ at which glass transition temperature $T_g$ is extracted.

## II. SIMULATION DETAILS

Our three dimensional cubic simulation box of side length $L$ with periodic boundary conditions applied in all directions is composed of $N = 1000$ (800 A and 200 B) particles with equal mass $M$. Density ($\rho$) is defined as $\rho = N/L^3$. Particles i and j, when their separation $r_{ij}$ is less than the potential cutoff $r_{ij}^c$, interact via a family of $(q, p)$ Lennard-Jones potentials having the following generalized form: [16,26]

$$V(r_{ij}) = \frac{\epsilon_{ij}}{q-p}\left[p\left(\frac{r_{ij}^0}{r_{ij}}\right)^q - q\left(\frac{r_{ij}^0}{r_{ij}}\right)^p\right] + f(r_{ij}), \quad (1)$$

where $r_{ij}^0 = 2^{1/6}\sigma_{ij}$ is the position of the minimum of $V(r_{ij})$ and the correction function $f(r_{ij})$ is added to satisfy continuity of $V(r_{ij})$ and its first derivative $V_0(r_{ij})$ at $r_{ij}^c$,[27] i.e. $V(r_{ij}^c) = V_0(r_{ij}^c) = 0$. Obviously, the steepness of repulsive part ($k$) can be tuned by various combinations of $(q, p)$, as shown in Fig. 1. We studied systems with six combinations of $(q, p)$ which are denoted in ascending order by steepness of repulsion as $k1(2,1)$, $k2(6,4)$, $k3(8,5)$, $k4(12,6)$, $k5(12,11)$ and $k6(18,16)$, respectively. For consistency and comparison with previous studies[16-19] in the same generalized Lennard-Jonnes potential systems, we will use denotations from $k1$ to $k6$ to characterize different steepness of the repulsive part of potential. Potential cutoffs $r_{ij}^c = 2.5\sigma_{ij}$ and $r_{ij}^c = 2^{1/6}\sigma_{ij}$ are chosen respectively for Lennard-Jones (**LJ**) system with both attraction and repulsion, and purely Repulsive Lennard-Jonnes (**RLJ**) system which is the same as the Weeks-Chandler-Andersen potential used in Refs. 24 and 25.

To suppress crystallization, we employ in our simulations the extensively used parameters proposed by Kob and Anderson in the standard Lennard-Jonnes potential[26] which is our model $k4(12,6)$: $\epsilon_{AB} = 1.5\epsilon_{AA}$, $\epsilon_{BB} = 0.5\epsilon_{AA}$, $\sigma_{AB} = 0.8\sigma_{AA}$, and $\sigma_{BB} = 0.88\sigma_{AA}$. And indeed no crystallization is observed in our simulation. The units for length, energy and mass scales are set to be $\sigma_{AA}$, $\epsilon_{AA}$ and $M$, respectively, and hence time is in units of $(M\sigma_{AA}^2/\epsilon_{AA})^{1/2}$. Temperature $T$ is expressed in units of $\epsilon_{AA}/k_B$ with $k_B$ the Boltzmann constant. We also performed simulations in $k4(12,6)$ model with other parameters as studied in Ref. 6 to make sure that our conclusions drawn from this study won't be affected by the choice of parameters for avoiding crystallization.

The fast inertial relaxation engine minimization algorithm[28] is employed to generate zero-temperature glasses (also termed inherent structures[9]) by quenching an initial high-temperature equilibrium state to its corresponding local potential energy minimum. The structure of inherent structures is measured by the pair distribution function of A particles, $g_{AA}(r) = \frac{L^3}{N_A^2}\langle\sum_i\sum_{j\neq i}\delta(r-r_{ij})\rangle$, where $N_A$ is the number of A particles, the sums are over all A particles and $\langle . \rangle$ denotes the average over different inherent structures.

For glass-forming liquids, molecular dynamics simulations are carried out in the canonical ensemble with a 4-variable Gear predictor-corrector integrator[27] used to propagate the motion of each particle. Data won't be collected until the system has been equilibrated for several structural relaxation time $\tau$ determined by the time when the self-intermediate scattering function of A particles, $F_s(q,t) = \frac{1}{N_A} < \sum_j \exp(i\vec{q} \cdot [\vec{r}_j(t) - \vec{r}_j(0)]) >$, with $\vec{r}_j(t)$ the location of particle j at time $t$, $\vec{q}$ chosen in the y direction with the amplitude approximately equal to the position of the first peak of the static structure factor,[27] the sum being taken over all A particles and $<.>$ denoting time average, decays to $e^{-1}$.[8] The glass transition temperature $T_g$ defined as $\tau(T_g) = 10^6$ in our study is extracted by fitting $\tau$ using Vogel-Fulcher-Tamman (VFT) function,[29] $\tau = \tau_0 \exp(\frac{C}{T-T_0})$, where $T_0$, $\tau_0$ and $C$, though having their separate physical significance,[19] are used here simply as fitting parameters. Fig. 2 illustrates how we determine $\tau$ from $F_s(q,t)$ and extract $T_g$ through VFT fit.

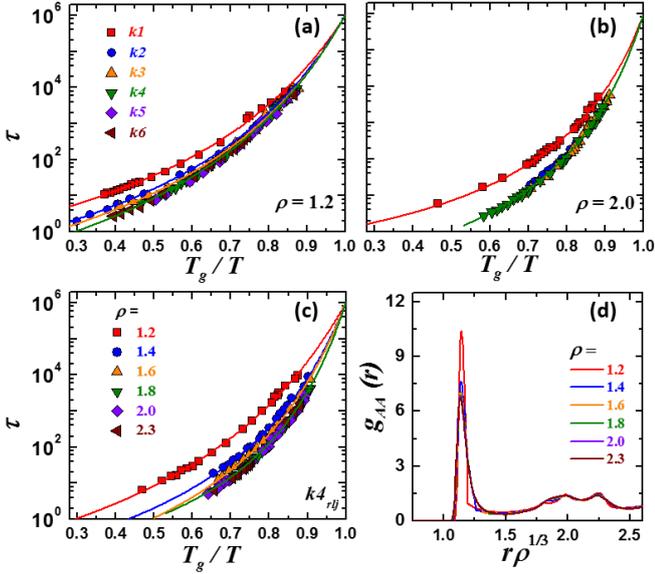

FIG. 3. (a), (b) and (c): Angell plot of relaxation time $\tau$ versus scaled reciprocal temperature $T_g/T$ in **RLJ** systems. Densities in panels (a) and (b) are 1.2 and 2.0, respectively, and symbols in panel (b) have the same implications as in (a). Data demonstrated in panel (c) is from model $k4_{rlj}$. Through data points in panels (a), (b) and (c) are VFT fit lines. (d): Scaling collapse of pair distribution functions of A particles $g_{AA}(r)$ at different densities in model $k4_{rlj}$.

### III. RESULTS AND DISCUSSION

Typical Angell plots of $\tau$ as a function of $T_g/T$ within **RLJ** system are shown in Fig. 3(a), (b) and (c) for various densities and all data can be well fitted by VFT function. Attention will be first focused on Fig. 3(a) which compares Angell plots between models from $k1_{rlj}$ to $k6_{rlj}$ at the widely used density,[26] $\rho = 1.2$. Seen clearly from $k1_{rlj}$ to $k4_{rlj}$, the curve becomes steeper with increasing steepness, indicating that fragility $m$ increases with enhanced steepness, i.e. softness makes strong glass formers, consistent with the colloidal experiment observation.[15] Then, an approximate collapse of different curves can be observed as the $k$ increases from $k4_{rlj}$ to $k6_{rlj}$ in Fig. 3(a), which suggests that there is an onset steepness $k_c$ above which $m$ becomes insensitive to the $k$. It is found that $k_c$ could be a function of $\rho$ and $k_c(\rho)$ decreases with larger density as evidenced in Fig. 3(b) for $\rho = 2.0$ where curves start to collapse at about $k2_{rlj}$. It demonstrates that the steepness dependence of fragility at fixed density exhibits disparate behaviours at different domains of steepness, and that softness makes strong glasses applies only in the context of a specific steepness range with $k$ weaker than $k_c(\rho)$. On the other hand, we also investigated the density-dependence of fragility for each steepness $k$. Fig. 3(c) shows the density dependence of fragility for $k4_{rlj}$, and the fragility increases with larger density, consistent with observations in other repulsive systems,[8,30] but attains constant above an onset density $\rho_c$, which, due to limited range of densities in previous studies, has never been observed before in **RLJ** systems. As shown in the inset to Fig. 1, the inter-particle distance $\langle r \rangle$ characterized by the first peak position of pair distribution functions, is a monotone decreasing function of density of this form, $\langle r \rangle \sim \rho^{-1/3}$. It means the region of steepness sampled by particles varies with $\rho$ for $k4_{rlj}$. Therefore, fragility's density dependence at fixed steepness model is, to a certain degree, analogous to its steepness dependence at fixed density. The onset density $\rho_c$ can be extracted for each fixed $k$ model and the invariant $m$ can be obtained for $\rho > \rho_c(k)$ cases.

Therefore, we enlarged our view and considered the entire $\rho - k$ field for **RLJ** systems. Fig. 4 shows the fragility $m$ as a two-dimensional function of $\rho$ and $k$. We found that the fragility in the $\rho - k$ field for **RLJ** systems can be divided into three regions with qualitatively distinct behaviours. The white solid line which is the upper boundary of Region **I** (denoted as **R$_I$**) is determined by the function of density-dependent onset steepness, $k = k_c(\rho)$. The conclusion about 'softness makes strong glasses' can only be achieved in **R$_I$** and fragility becomes insensitive to the $k$ for models in Region **II** (denoted as **R$_{II}$**). The fragility in **R$_{II}$** can only be tuned by $\rho$, which suggests the profile of potential energy landscape (PEL)[9] of systems belonging to **R$_{II}$** is mainly dominated by particle density. The lower boundary of Region **III** (denoted as **R$_{III}$**) is described by the function of steepness-dependent onset density, $\rho = \rho_c(k)$ and the models in **R$_{III}$** present uncontrollable constant fragility. It implies that the systems in **R$_{III}$** presenting identical dynamic evolution behaviours near $T_g$ may provide us a good platform to understand the glass transition.

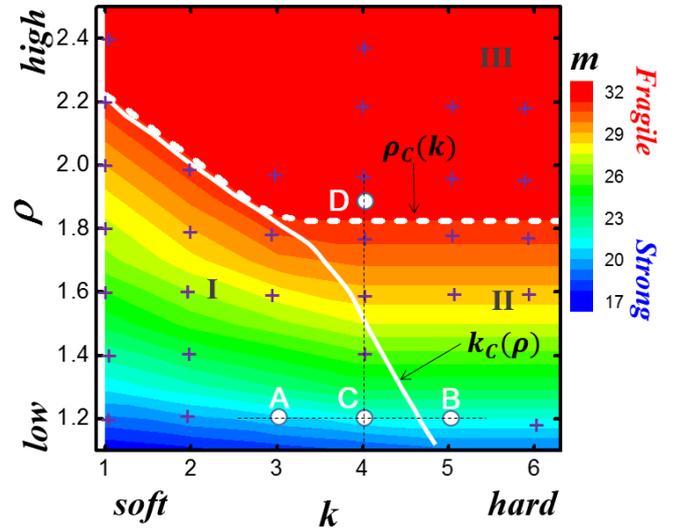

FIG. 4. Schematic contour plot of fragility ($m$) as a function of density ($\rho$) for different steepness models ($k$) in **RLJ** systems. The crosses in the panel indicate all the data points we have obtained in our study and the circles mark the locations of the four models which will be discussed in detail in Fig. 5. More detailed description of the plot can be found in the text.

For a serial of systems with same $k$ but different $\rho$, the density-dependent dynamic behaviors near $T_g$ can be understood by the structural properties of related zero-temperature glassy states. The pair distribution functions of A particles $g_{AA}(r)$ against $r\rho^{1/3}$ are shown in Fig. 3(d) for $k4_{rlj}$. It presents nice scaling collapse above a crossover density which almost coincides with $\rho_c(k)$ ($\rho_c(k) = 1.8$ for $k4_{rlj}$ as shown in Fig. 3(c)). Similar scaling collapse has recently been observed in LJ glasses[24] and liquids[31]. We find the similarities of structures of inherent structures above the steepness-dependent crossover density can be observed in all the six steepness models in our study. Even for the softest model $k1_{rlj}$, the heights of the first peak of $g_{AA}(r)$ almost don't vary with densities for $\rho > 2.0$ (not shown here). Therefore, for one certain $k$, the invariant fragility due to the high density effect can be explained by the isomorph of inherent structures. However, it should be mentioned that neither isomorphic structures nor similarities of structures can be observed to explain the hard steepness effect at fixed density. It suggests that the two-point-correlation function $g(r)$ can't provide a full understanding of dynamic behaviors and hence high-order correlation information should be included.

So far we are concerned with steepness dependence of fragility in purely repulsive systems. Generally, attractive interactions are usually treated as perturbation in the theory of liquids.[32] However, it has been recently demonstrated that glass formers with attraction exhibit more sluggish[25] and heterogeneous[23] dynamics than those with pure repulsion, and hence attraction doesn't act as perturbation. Though the non-perturbative effect of attraction on some quantities[23-25] under certain conditions[24] has been put forward for years, to our knowledge, no attention has yet been focused on probing the attraction dependence of fragility. In order to probe whether and how attraction affects fragility, we chose four **RLJ** models (marked as **A**, **B**, **C**, and **D**) which are located on different regions as shown in Fig. 4, and introduced the attraction into the atomic interaction. The calculated Angell plots are shown in Fig. 5 (a) and (b), and the data of related **RLJ** systems are also present for comparison. The divergences between the plots of the models (**A**, **B** and **C**) with and without attraction indicate that the attraction in the systems belonging to **R$_I$** can't be treated as perturbation and makes them more fragile than respective **RLJ** models. As shown in Fig. 5(b), the collapse of the two curves of model **D** (with and without attraction, respectively) indicates that the attraction effect in **R$_{III}$** can be avoided and we can employ the **RLJ** model to investigate the dynamic behaviors of the respective **LJ** model in this region. For extracting the contribution of the attraction part quantitatively, we calculate the fragility $m$ for systems $k_1 \sim k_6$ at $\rho = 1.2$, respectively. The results are shown in Fig. 5(c). In **RLJ** systems, we can make the same conclusion in **R$_I$** as in the colloidal experiment[15] that softness makes strong glasses. In **LJ** systems, the contribution of attraction produces a reversal correlation between $k$ and $m$, implying that hardness makes strong glass, which is consistent with recent studies.[16-19] However, the attraction makes huge contribution in **R_I**, so we can't simply attribute the different dynamic behaviors between **LJ** models to the steepness of the repulsive part of interaction. The influence of attraction weakened as $k$ enhanced (or $\rho$ increased), which suggests that the profile of PEL is controlled by repulsive part as the repulsive force becomes the dominated interaction.

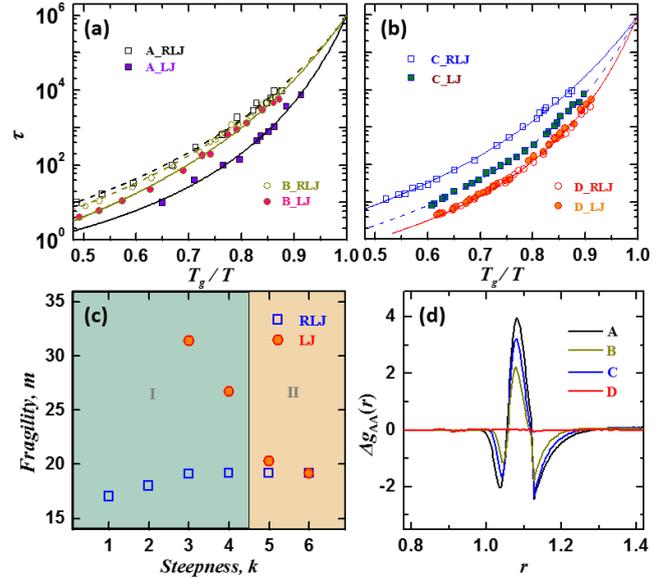

FIG. 5. (a) and (b): Angell plot of relaxation time $\tau$ versus scaled reciprocal temperature $T_g/T$ for models $(\rho, k) = (1.2, k3)$, $(1.2, k5)$, $(1.2, k4)$ and $(1.9, k4)$ which are marked as **A**, **B**, **C** and **D** in Fig. 4, respectively. Through data points in panels (a) and (b) are VFT fit lines. (c): Steepness $k$ dependence of fragility index $m$ in **RLJ** and **LJ** systems at fixed density, $\rho = 1.2$. The two regions **I** and **II** with different colors are subsets of **R$_I$** and **R$_{II}$** in Fig. 4, respectively. (d): Comparison of $\Delta g_{AA}(r)$ as a function of $r$ between models **A**, **B**, **C** and **D**.

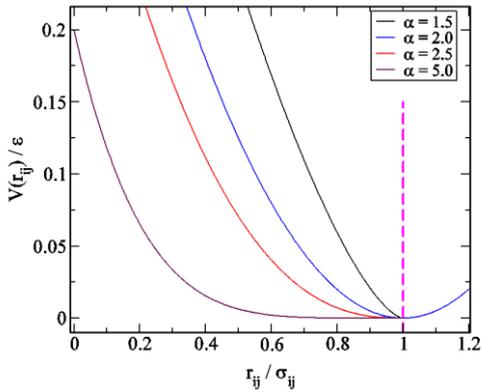

FIG. 6. Schematic illustration of the scaled pair core-softened potential $V(r_{ij})/\epsilon$ for $\alpha$ = 1.5, 2.0, 2.5 and 5.0, respectively. The dashed line marks the position of $r_{ij}/\sigma_{ij} = 1$ which separates attractive from repulsive part of the interaction potential.

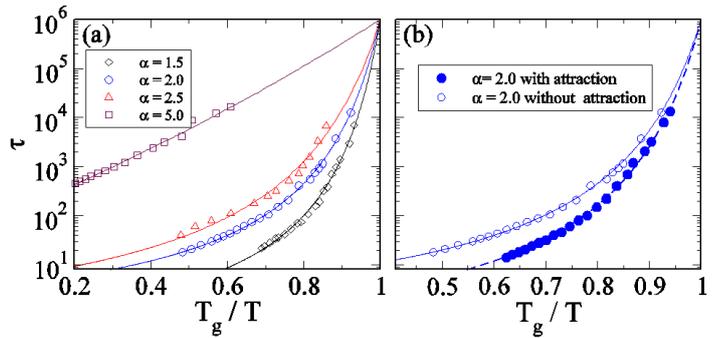

FIG. 7. Angell plot of relaxation time $\tau$ versus scaled reciprocal temperature $T_g/T$ at volume fraction $\phi$ = 0.8. (a): Comparison between four purely repulsive core-softened systems ( $r_c$ = 1.0 ) in ascending order by softness of repulsion with $\alpha$ = 1.5, 2.0, 2.5 and 5.0. (b): Comparison between systems for $\alpha$ = 2.0 with attraction ($r_c$ = 1.2) and pure repulsion ($r_c$ = 1.0). Through data points in panels (a) and (b) are VFT fit lines.

It has been shown that structures of inherent structures exhibit difference between **RLJ** and **LJ** systems at fixed density when attraction in **LJ** system doesn't act as perturbation.[24] We find the difference level of attraction influence on fragility can also be estimated by the attraction effect within the structural properties of zero-temperature glasses. Fig. 5 (d) presents the structural difference between **RLJ** and **LJ** systems, which is defined by $\Delta g_{AA}(r) = g_{AA,rlj}(r) − g_{AA,lj}(r)$, to quantify the extent of attraction influence in **LJ** system. Obviously, the attraction is perturbative in **LJ** system when $\Delta g_{AA}(r) = 0$ for any distance $r$ (such as in model **D**), and bigger $\Delta g_{AA}(r)$ reflects more important role of attraction (such as in models **A**, **B** and **C**). Thus, the extent of attraction influence on structure correlates well with that of fragility difference between **RLJ** and **LJ** systems in models **A**, **B**, **C**, and **D**, respectively, as shown in Fig. 5 (a), (b) and (d).

## IV. FURTHER DISCUSSION IN CORE-SOFTENED POTENTIAL SYSTEMS

Meanwhile, to further testify our conclusions drawn from Lennard-Jonnes systems, we extend our research on softness dependence of fragility to the widely used core-softened potential system mimicking colloids with a binary (50 : 50) mixture of N = 1000 frictionless spheres having the same mass.[8,33,36] By convention, the diameter ratio of the mixture is set to be 1.4 to avoid crystallization , and the detailed parameters related to units setting for core-softened potential can be found in Ref. [8]. The interaction in core-softened potential between particles i and j is $V(r_{ij}) = \frac{\epsilon}{\alpha}\left(1 - \frac{r_{ij}}{\sigma_{ij}}\right)^{\alpha}$ when their separation $r_{ij}$ is smaller than $r_c\sigma_{ij}$ with $\sigma_{ij}$ being the sum of their radii, $\epsilon$ the energy scale and $r_c$ the potential cutoff, and zero otherwise. $r_c$ is chosen to be 1.0 to mimic purely repulsive systems and we set $r_c$ to be a value larger than 1.0 to introduce nonperturbative attraction, e.g. $r_c$ = 1.2 in system with $\alpha$ = 2.0 in this study. We study four systems with $\alpha$ = 1.5, 2.0, 2.5 and 5.0, respectively, at the same volume fraction $\phi$ defined as $\phi = (\sum_{i=1}^{N} \sigma_i^3)/L^3$. The schematic plots of the four core-softened potentials with the absence and presence of attraction are shown in Fig. 6. Seen from Fig. 6 the curves get steeper with decreasing $\alpha$. Thus, the parameter $\alpha$ can be used to tune the softness of interaction, i.e. larger value of $\alpha$ is indicative of softer interaction.

Figure 7 (a) compares Angell plot between the four purely repulsive core-softened potential systems with different softness of repulsion. The curve becomes more flat with larger value of $\alpha$ (softer repulsion), suggesting that softness makes strong glass formers, which is the major conclusion about the softness dependence of fragility in previous experimental[15] and our theoretical studies. It should be noted that we also get the similar conclusion at other volume fractions above or below jamming transition point[6] in purely repulsive core-softened potential systems.

We can't get models with arbitrary small softness in the generalized Lennard-Jones potential as long as we keep the exponents in the potential positive to make the potential possess physical meaning. It is found that the variance of the fragility in Fig. 3 (a) or (b) is small compared with the experiment in Ref. [15]. However, the core-softened potential whose softness of repulsion can be tuned to be arbitrarily small, can approximate some experimental colloidal systems,[33] which is also one of our purposes to include the discussion of the core-softened potential systems here. In core-softened systems, we can see nearly as large variance of fragility in Fig. 7 (a) as observed in the experimental study,[15] though the interaction potential between colloidal particles in Ref. [15] are not the same as our core-softened potential. It should be mentioned that the steepness of core-softened potential can't be tuned to be arbitrary hardness due to its expression, and hence we can't observe the invariant fragility (only existing at very hard steepness systems based on the study from repulsive Lennard-Jonnes systems) at fixed volume fraction within the steepness range we can get access to in repulsive core-softened systems. Therefore, it seems that repulsive core-softened and repulsive Lennard-Jonnes potentials are complementary in studying the softness dependence of fragility.

Figure 7 (b) compares Angell plot between the core-softened potential ($\alpha = 2.0$) systems with the absence and presence of attraction. As demonstrated in it, the attraction within core-softened potential systems induces enhancement of fragility, which suggests that attraction makes fragile glasses is the general conclusion applying to different potential systems. Though larger $m$ in system with attraction may be related to the finding that attractive system is more heterogeneous in dynamics than its counterpart without attraction,[23,34] more convincingly microscopic explanation needs to be further explored.

## V. CONCLUSIONS

In this study, through MD simulations and including the effect of attraction, we provided a qualitative picture of how interaction potential and density determine fragility of glass forming liquids, and unified the long-standing inconsistence pertaining to the repulsive softness dependence of fragility. In purely repulsive Lennard-Jonnes systems, the fragility in the $\rho - k$ field can be divided into three regions qualitatively: $R_I$ which can be described by 'softness makes strong glasses', $R_{II}$ where fragility can only be tuned by $\rho$, and $R_{III}$ with constant fragility. It suggests the density plays an unexpected role as we discuss the repulsive steepness dependence of fragility. The level of the influence of attraction on fragility can be estimated from the structure information of related zero-temperature glasses, and a glass former will be tuned more fragile if nonperturbative attraction is added, especially in $R_I$. However, due to the diverse correlations between the fragility, density and atomic interaction within different regions, to simply claim either repulsion or attraction determines fragility is not all-inclusive and sometimes may give rise to inconclusive studies.

Our findings in the course of this study may point out directions for further research. Our simulations are mainly conducted at densities no less than $\rho = 1.2$ within Kob-Anderson parameters.[26] We will enter at lower densities the scenario of jamming transition or hard-sphere glass transition in repulsive systems[3,6,35,36] and the territory of inhomogeneous liquids or gels with formation of cavitation in attractive systems,[18,19,37] which are beyond the scope of this study. It will also be interesting to investigate systematically how the varying repulsive steepness of interaction affects fragility when the strength of attraction is held fixed as studied in Ref. 38, and how the varying length and strength of attraction influence fragility with the repulsive steepness of interaction unchanged as done in Ref. 39. We also note that the scaling collapse of $g(r)$ measured at three quite high pressures, 2.5, 18.2 and 31.4 GPa, is presented in a recent experiment study[40] on $Pd_{81}Si_{19}$ glassy alloy whose pairwise interaction approximates Lennard-Jonnes potential. Therefore, if possible, the high-pressure experiment may be a good choice to verify our finding of the invariant fragility above $\rho_c$ for specific interatomic potential. Furthermore, density, repulsion and attraction, when mixed together, can make a variety of changes of fragility, which will be an optional platform to test the generality of the known correlations between fragility and other parameters, such as Poisson's ratio[14] and anharmonicity.[16,17]

## ACKNOWLEDGEMENTS

We thank N. Xu and L. M. Xu for useful discussions. We also acknowledge the computational support from the Beijing Computational Science Research Center (CSRC). The work was supported by the NSF of China (Grant No. 51571111) and the MOST 973 Program (No. 2015CB856800).